\documentclass[prc,superscriptaddress,twocolumn,showpacs,preprintnumbers,amsmath,amssymb,floatfix,dvipdfmx]{revtex4}

\usepackage{amsmath}
\usepackage{amssymb}
\usepackage{mathrsfs}
\usepackage{bm}
\usepackage{color}
\usepackage{graphicx}
\usepackage{braket}
\usepackage{ulem}

\makeatletter
\let\MYcaption\@makecaption
\makeatother
\usepackage{subcaption}
\makeatletter
\let\@makecaption\MYcaption
\makeatother

\makeatletter
\newcommand{\figcaption}[1]{\def\@captype{figure}\caption{#1}}
\newcommand{\tblcaption}[1]{\def\@captype{table}\caption{#1}}
\makeatother

\begin{document}

\preprint{NITEP 96}
\title{
Investigation of multi-step effects for proton inelastic scattering to the $2^{+}_{1}$ state in $^6$He
}

\author{Shoya Ogawa}
\email[]{s-ogawa@phys.kyushu-u.ac.jp}
\author{Takuma Matsumoto}
\email[]{matsumoto@phys.kyushu-u.ac.jp}
\affiliation{Department of Physics, Kyushu University, Fukuoka 819-0395, Japan}

\author{Yoshiko Kanada-En'yo}
\email[]{yenyo@ruby.scphys.kyoto-u.ac.jp}
\affiliation{Department of Physics, Kyoto University, Kyoto 606-8502, Japan}

\author{Kazuyuki Ogata}
\email[]{kazuyuki@rcnp.osaka-u.ac.jp}
\affiliation{Research Center for Nuclear Physics (RCNP), Osaka University,
 Ibaraki 567-0047, Japan}
\affiliation{Department of Physics, Osaka City University, Osaka 558-8585, Japan}
\affiliation{Nambu Yoichiro Institute of Theoretical and Experimental Physics (NITEP), 
Osaka City University, Osaka 558-8585, Japan}

\date{\today}

\begin{abstract}
 Multi-step effects between bound, resonant, and non-resonant states
 have been investigated by the continuum-discretized coupled-channels method
 (CDCC). In the CDCC, a resonant state is treated as multiple states
 fragmented in a resonance  energy region, although it is described as a
 single state in usual coupled-channel calculations. For such the
 fragmented resonant states, 
 one-step and multi-step contributions to the cross sections should be
 carefully discussed because the cross sections obtained by the one-step
 calculation depend on the number of those states, which corresponds to
 the size of the model space. To clarify the role of the multi-step
 effects, we propose the one-step calculation without model-space
 dependence for the fragmented resonant states. Furthermore, we also
 discuss the multi-step effects between the ground, $2^+_1$ resonant,
 and non-resonant states in  $^6$He for proton inelastic scattering. 
\end{abstract}

\maketitle

%
\section{Introduction}
\label{sec:intro}

Resonances are metastable states that appear beyond the particle decay threshold
in numerous quantum systems.
In nuclei, there exist various resonances,
such as single-particle resonances, giant resonances, and cluster resonances,
that reflect various nucleon correlations. The investigation of such resonances
has attracted substantial attention.
Recently, resonances of nuclei near or beyond the neutron dripline have been
investigated through radioactive ion-beam experiments
\cite{Spy12,Kan15,Kon16,Tan17,Sun21}.
To elucidate the properties of resonances,
their transition cross sections can be experimentally measured
via inelastic scattering and/or transfer reactions.
Therefore, it is highly desirable to analyze the experimental data
by precise theoretical calculations to extract the properties of resonances.

Coupled-channels (CC) calculations provide a reliable method to describe 
the transition to excited bound and resonant states and have been
successfully applied to the analysis of various reactions
\cite{Kho08,Fur13,Tak10,Kan19-1,Kan19-2,Kan19-3,Kan20-1}. %
In CC calculations, higher-order coupling effects between the states considered
in the scattering are included.
As an alternative to CC calculations,
approximate calculations considering only the one-step process have been frequently applied
to the analysis of inelastic scattering~\cite{Lag01,Ste02,Kho05,Fal13,Tan17,Gas18}.
These approximations are referred to as one-step calculations,
and the difference between the results of CC and one-step calculations represents 
the multi-step effects.
At intermediate and high incident energies,
the multi-step effects become small and one-step calculations work well in general.

Recently,
proton inelastic scattering to the $2^+_1$ state of $^6$He at 41 MeV/nucleon~\cite{Lag01}
was analyzed by two methods,
namely, the microscopic CC method based on antisymmetrized molecular dynamics 
(AMD)~\cite{Kan20-2},
which is referred to as AMD-MCC,
and the continuum-discretized coupled-channels method (CDCC)~\cite{Oga20}.
In the AMD-MCC,
the $2^+_1$ state is represented by a bound-state approximation,
while the coupling potential between the ground and $2^+_1$ states in $^6$He is calculated
by the microscopic folding model with the Melbourne $g$
matrix~\cite{Amo00,Lag01,Ste02,Kar07,Min10,Toy13,Ega14,Min16,Min17}.
The AMD-MCC has been successfully applied to the analysis of various cases of inelastic
scattering~\cite{Kan19-1,Kan19-2,Kan19-3,Kan20-1,Kan20-2,Kan20-3,Kan20-4,Kan20-5,Kan20-6,Kan21-1}.
On the other hand,
the CDCC takes into account coupling effects to not only the $2^+_1$ state but also 
non-resonant states,
which are represented by a finite number of discretized states.
The CDCC has also been successfully used to describe reactions involving
unstable nuclei~\cite{Yah12,Mat04,Mat06,Mat19}.
Thus, although both methods reproduce the inelastic data reasonably well,
they afford different views of the multi-step effects for the $2^+_1$ state.

In the AMD-MCC,
the inelastic cross section obtained from the CC calculations is in good agreement
with that derived from one-step calculations.
This result demonstrates that the multi-step effects between the ground and $2^+_1$ states 
are small.
On the other hand,
the cross section obtained from the CDCC is not consistent with that provided by 
the one-step calculations,
indicating that the multi-step effects make a significant contribution to the inelastic 
cross section.

One of reasons for this discrepancy concerning the significance of
multi-step effects 
is considered to be the influence of non-resonant states,
which are taken into account in the CDCC but not in the AMD-MCC. 
In addition, the resonant state is fragmented into multiple states in the CDCC.
The one-step calculation in the CDCC neglects the multi-step effects between the
fragmented resonant states, which are significant because the resonant state can be
regarded as a single state.
Furthermore
it is known that the cross sections to the fragmented states in the one-step calculations
are dependent on the number of these states, which corresponds to the size of the model space.
To clarify the coupling effects to the non-resonant states,
detailed examination of the treatment of the fragmented resonant states is required.

In this paper,
we examine the contributions of fragmented resonant states to the inelastic cross section
and attempt to clarify the multi-step effects between the ground and resonant states.
To this end, we propose a new treatment for the fragmented states without the model-space 
dependence.
To confirm the validity of this approach, we first analyze the three-body reaction
of the $^6$Li + $^{40}$Ca system, where $^6$Li is described as a $d$ + $\alpha$ two-body 
system,
because it is easy to adjust the size of the model space.
Finally, we discuss the multi-step effects in the proton inelastic
scattering of $^6$He. 

This paper is organized as follows. In Sec. II, we describe the
theoretical framework. In Sec. III, we present and discuss the numerical results.
Finally, in Sec. IV, we provide a summary of the key findings.

%
\section{Formalism}
\label{sec:formalism}

In the present work,
we consider a projectile breakup reaction in which the projectile has
one bound state and one resonant state.
In the CDCC, the reaction is assumed to take place in a model space ${\cal P}$ defined by
\begin{eqnarray}
 \mathcal{P}=
  |\Phi_0\rangle\langle \Phi_0|+
  \sum_{\gamma=1}^{N} \ket{\Phi_{\gamma}}\bra{\Phi_{\gamma}}   
  \equiv{\cal P}_0+\sum_{\gamma}{\cal P}_\gamma,
\end{eqnarray}
where $\Phi_{0}$ and $\Phi_{\gamma}$ are the wavefunctions of the ground state and
the discretized continuum states with quantum number $\gamma$, respectively.
${\cal P}_0$ and ${\cal P}_\gamma$ are projection operators on $\Phi_0$ and $\Phi_\gamma$, 
respectively.
It should be noted here that $\Phi_\gamma$ includes the fragmented resonant states.
In the present analysis,
we adopt the pseudo-state discretization with the Gaussian expansion method
to obtain a set of $\{\Phi_\gamma\}$.

In the space $\mathcal{P}$, the Schr\"{o}dinger equation for the scattering can be expressed 
as
\begin{eqnarray}
\mathcal{P} ( K+U+h - E ) \mathcal{P} \Psi = 0 ,
\label{sch-eq}
\end{eqnarray}
where $K$ and $U$ denote the kinetic energy and potential between the projectile and target,
respectively.
The internal Hamiltonian of the projectile $h$ satisfies
\begin{eqnarray}
 \varepsilon_0=\langle \Phi_0|h|\Phi_0\rangle,\quad
 \varepsilon_\gamma=\langle \Phi_\gamma|h|\Phi_\gamma\rangle. 
\end{eqnarray}
Multiplying ${\cal P}_0$ and ${\cal P}_\gamma$ from the left side in Eq.~\eqref{sch-eq} leads
to the CDCC equation
\begin{eqnarray}
 \left[
  K + {\cal P}_0 U{\cal P}_0 - (E - \varepsilon_{0})
 \right]
 {\cal P}_0\Psi
 =
 - \sum_{\gamma'\ne 0}
 {\cal P}_{0}U{\cal P}_{\gamma'}\Psi,
\nonumber \\
 \left[
  K + {\cal P}_\gamma U{\cal P}_\gamma - (E - \varepsilon_{\gamma})
 \right]
 {\cal P}_\gamma\Psi
 =
 - \sum_{\gamma'\ne \gamma}
 {\cal P}_{\gamma}U{\cal P}_{\gamma'}\Psi.
 \label{eq:cc-eq}
\end{eqnarray}
In the CDCC,
${\cal P}_0\Psi$ and ${\cal P}_\gamma\Psi$ can be solved under appropriate 
boundary conditions,
the details of which can be found in Ref.~\cite{Yah12}.

In contrast to Eq. \eqref{eq:cc-eq} of the CDCC calculation,
the equation of the one-step calculation is given as
\begin{eqnarray}
 \label{eq:1step-eq}
 \left[
  K + {\cal P}_0U{\cal P}_{0} - (E - \varepsilon_{0})
 \right]
 {\cal P}_0\Psi
 &=&
 0 , 
 \nonumber \\
 \left[
  K + {\cal P}_\gamma U{\cal P}_\gamma - (E - \varepsilon_{\gamma})
 \right]
 {\cal P}_\gamma\Psi
 &=&
 -  {\cal P}_\gamma U{\cal P}_0\Psi .
\end{eqnarray}
The first line of Eq.~\eqref{eq:1step-eq} is not a coupling equation,
and coupling effects from other states are neglected.
In the second line of Eq.~\eqref{eq:1step-eq},
only the one-step coupling effect from ${\cal P}_0\Psi$ is taken into account.
It should be noted that the coupling effects between fragmented resonant states
are also neglected in Eq.~\eqref{eq:1step-eq}.

In the CDCC, 
we assume that the resonant model space $\mathcal{P}_{\rm R}$
can be described by the sum of the fragmented resonant states as 
\begin{eqnarray}
 {\cal P}_{\rm R}
  \equiv
  \sum_{\gamma_{\rm R}}|
  \Phi_{\gamma_{\rm R}}\rangle\langle\Phi_{\gamma_{\rm R}}|\equiv
  \sum_{\gamma_{\rm R}}
  {\cal P}_{\gamma_{\rm R}},
  \label{P_r}
\end{eqnarray}
where $\gamma_{\rm R}$ means the quantum number of the fragmented resonant state
in $\gamma$.
Considering the one-step transition to $\mathcal{P}_{\rm R}$,
the equation to be solved is represented as
\begin{eqnarray}
 \left[
  K + {\cal P}_{\rm R}U{\cal P}_{\rm R} - (E - h)
 \right]
 {\cal P}_{\rm R}\Psi
 &=&
 - {\cal P}_{\rm R}U{\cal P}_0\Psi.
 \label{res1-eq}
\end{eqnarray}
Multiplying ${\cal P}_{\gamma_{\rm R}}$ from the left-hand side of Eq.~\eqref{res1-eq}, 
the coupling equation for $\mathcal{P}_{\gamma_{\rm R}}$ is obtained as
\begin{eqnarray}
 \label{eq:1step-eq-res2}
&& \left[
  K + {\cal P}_{\gamma_{\rm R}}U{\cal P}_{\gamma_{\rm R}} - (E -
  \varepsilon_{\gamma_{\rm R}})
 \right]{\cal P}_{\gamma_{\rm R}}\Psi
\nonumber\\
 &&\hspace{1.5cm}=
 - {\cal P}_{\gamma_{\rm R}}U{\cal P}_0\Psi
 - \sum_{\gamma_{\rm R}'\ne \gamma_{\rm R}}{\cal P}_{\gamma_{\rm R}}U{\cal
 P}_{\gamma_{\rm R}'}\Psi.
\end{eqnarray}
In Eq.~\eqref{eq:1step-eq-res2},
one can see that only coupling effects between the fragmented resonant states are taken 
into account. 
In this work,
calculations with Eqs.~\eqref{eq:1step-eq} and \eqref{eq:1step-eq-res2} are referred to as
the {\it conventional one-step calculation} and {\it resonant one-step calculation}, 
respectively.
It should be noted here that the discretization of the resonant state does not affect 
the CDCC equation of Eq.~\eqref{eq:cc-eq}.

In practice,
we cannot define ${\cal P}_{\rm R}$ exactly in the CDCC calculation 
because the fragmented resonant states include not only the resonant component
but also non-resonant components.
Thus, we select the fragmented resonant states by determining the energy range in the breakup
cross section as described in Sec.~III.
In the present analysis, the inelastic cross section describing the transition from 
the ground state to the resonant state is defined by
\begin{eqnarray}
 \label{eq:inela}
  {\frac{d\sigma_{inel.}}{d\Omega}}
  =
  \sum_{\gamma_{\rm R}} 
  {\frac{d\sigma_{\gamma_{\rm R}}}{d\Omega}},
\end{eqnarray}
where $d\sigma_{\gamma_{\rm R}}/d\Omega$ represents the cross sections to the fragmented 
resonant states.

\section{Results and Discussion}
\label{sec:results}

\subsection{$^6$Li scattering} 
First, we investigate the model-space dependence of the one-step calculation
through the analysis of a case of three-body scattering,
in which the projectile is described by a two-body model.
To this end, we analyze the $^6$Li + $^{40}$Ca reaction at $E$ = 26 MeV/nucleon,
which was reported in Ref.~\cite{Mat03}.
The scattering is described by the $d$ + $\alpha$ + $^{40}$Ca three-body model.
The model Hamiltonian $H$ is the same as that used in Ref.~\cite{Mat03}.
In this model, the spin of the deuteron in $^6$Li is neglected,
the ground state of $^6$Li has the total spin-parity $I^\pi=0^+$,
and there is one resonant state in $I^\pi=2^+$.
The resonant energy and decay width are 2.75 MeV and 0.2 MeV, respectively.

\begin{figure}[tb]
 \centering
 \includegraphics[scale=0.45]{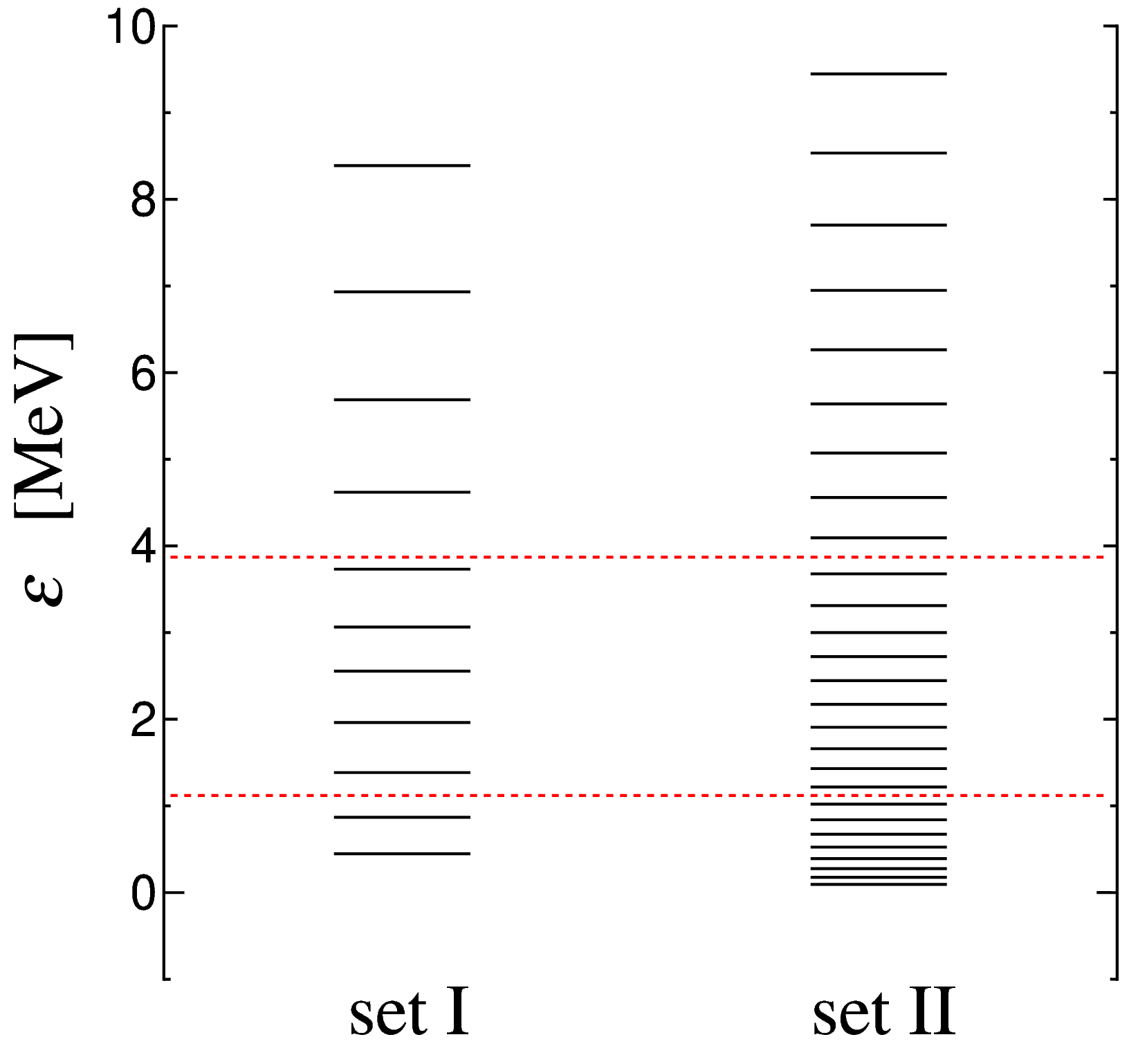}
 \caption{
 Pseudo states of $^6$Li for $I^{\pi}$ = $2^+$ calculated with sets I and II.
 On the vertical axis, 0 MeV corresponds to the $d$ + $\alpha$ threshold.
 States in the area between the two lines represent fragmented resonant states. 
 }
 \label{fig:6Li-res}
\end{figure}

\begin{figure}[tb]
 \centering
 \includegraphics[scale=0.80]{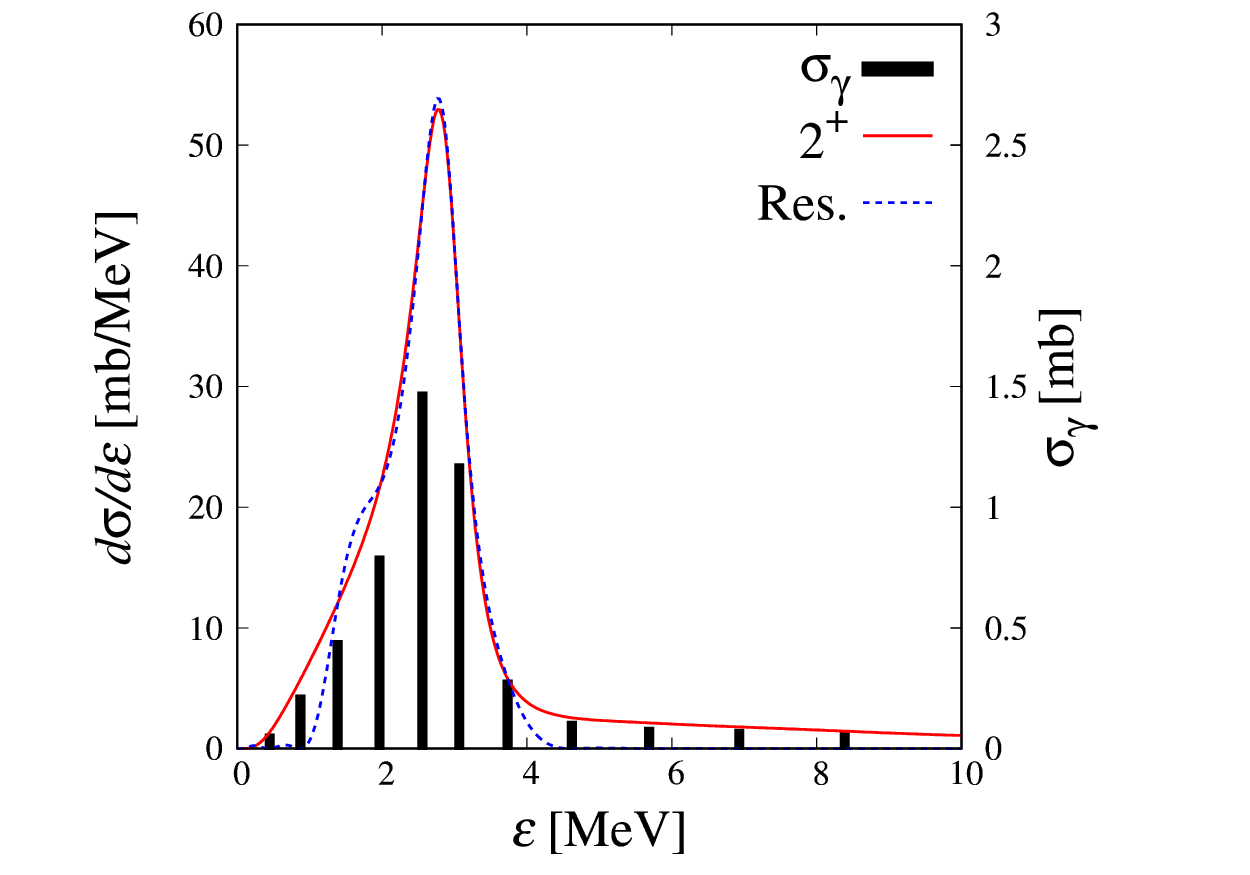}
 \caption{
 Breakup cross section for $^6$Li + $^{40}$Ca scattering calculated with set I.
 The bars represent the breakup cross sections to discretized states,
 and the absolute values are shown on the right vertical axis.
 }
 \label{fig:6Li40Ca-dbu}
\end{figure}

To study the model-space dependence,
we prepare two sets of the discretized states for $I^\pi=2^+$ of $^6$Li as set I and set II.
The number of discretized states in set II is larger than that in set I,
as shown in Fig.~\ref{fig:6Li-res}.
Here, set I is calculated with the same parameters as in Ref.~\cite{Mat03}
and gives good convergence of the CDCC.
In the pseudo state discretization,
the resonant and non-resonant components of a two-body structure can be rather clearly 
distinguished because the breakup cross sections to discretized states calculated
with set I smoothly change with the internal energy of $^6$Li, as shown by histograms
in Fig.~\ref{fig:6Li40Ca-dbu}.
As the fragmented resonant states,
we select states within the range of $1.1$ MeV $\leq \varepsilon\leq$ $3.9$ MeV,
which corresponds to the region between the two dashed lines in Fig~\ref{fig:6Li-res}.
To clarify the validity of the selection,
we calculate the breakup cross section as a function of the internal energy $\varepsilon$
of $^6$Li.
In the CDCC, the breakup cross section is described as,
\begin{eqnarray}
\label{eq:bxs}
 \frac{d\sigma}{d\varepsilon}
  =
  \left|
   \sum_{\gamma} f_{\gamma}(\varepsilon) T_{\gamma}
  \right|^{2} .
\end{eqnarray}
$T_{\gamma}$ is a discretized $T$-matrix obtained by the CDCC.
The histogram values describe $\sigma_{\gamma} = |T_{\gamma}|^{2}$.
$f_{\gamma}(\varepsilon)$ represents the smoothing function and
the details can be found in Ref.~\cite{Mat03}.
In Fig. \ref{fig:6Li40Ca-dbu}, 
the solid line shows the calculated breakup cross section to $I^\pi=2^+$ continuum states
with set I, where all discretized states are taken in the summation of Eq. \eqref{eq:bxs}.
Meanwhile, the dotted line represents the result taking only the fragmented resonant states.
One can see that the solid line is in good agreement with the dotted line around 
the resonance energy.
In the case of set II, we performed the same analysis and found that the energy range
is the same as one of set I.
Thus we conclude that the states within the range of 
$1.1$ MeV $\leq \varepsilon\leq$ $3.9$ MeV correspond to the fragmented resonant states.
The number of the fragmented resonant states in set~I (set~II) is five (ten).
We discuss the behavior of wavefunctions of discretized states in Appendix.
\begin{figure}[t]
 \centering  
 \includegraphics[scale=0.80]{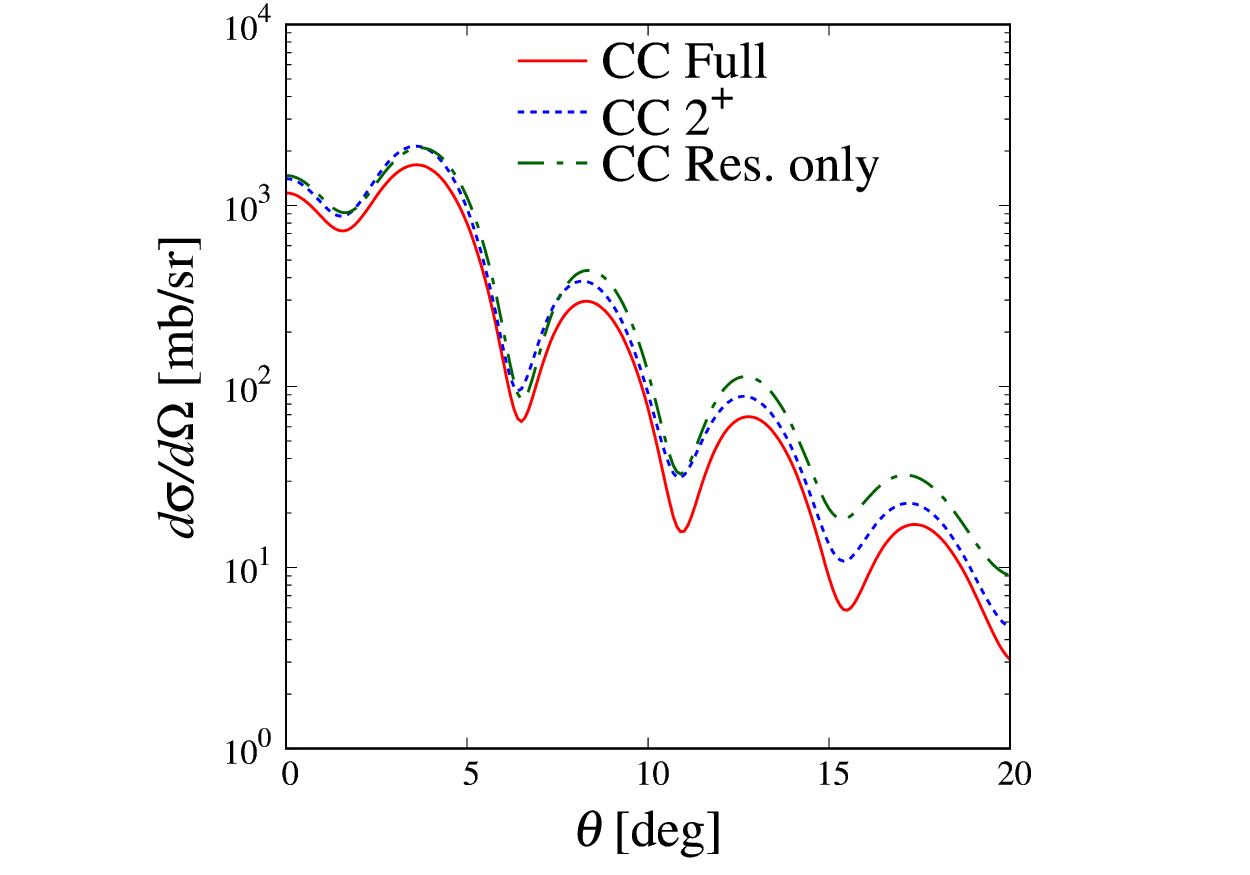}
 \caption{
 Inelastic cross sections for the $^6$Li + $^{40}$Ca reaction at 
 $E$ = 26 MeV/nucleon obtained from the CDCC.
 The pseudo states considered in the results are calculated with parameter set I 
 for Gaussian bases.
 }
 \label{fig:6Li-inela}
\end{figure}

In Fig.~\ref{fig:6Li-inela}, we show the inelastic cross sections 
to the fragmented resonant $2^+_1$states calculated
by the CDCC using Eq.~\eqref{eq:inela} with set I.
It should be noted that if set II is used in the calculations, we obtain
the same results. The solid line shows the result of the CDCC with
coupling effects of all states for $I^\pi=0^+$ and $2^+$. The dotted and
dot-dashed lines indicate the results of the CDCC with all $2^+$ states
including non-resonant states and only the fragmented resonant states,
respectively. The difference between the solid and dotted lines
represents the coupling effects for $I^\pi=0^+$, and it reduces the
cross section by approximately 20--30\%. In contrast, the difference
between the dotted and dot-dashed lines, which corresponds to the
coupling effects of non-resonant states in $I^\pi=2^+$, is negligible. 

\begin{figure}[t]
 \includegraphics[scale=0.80]{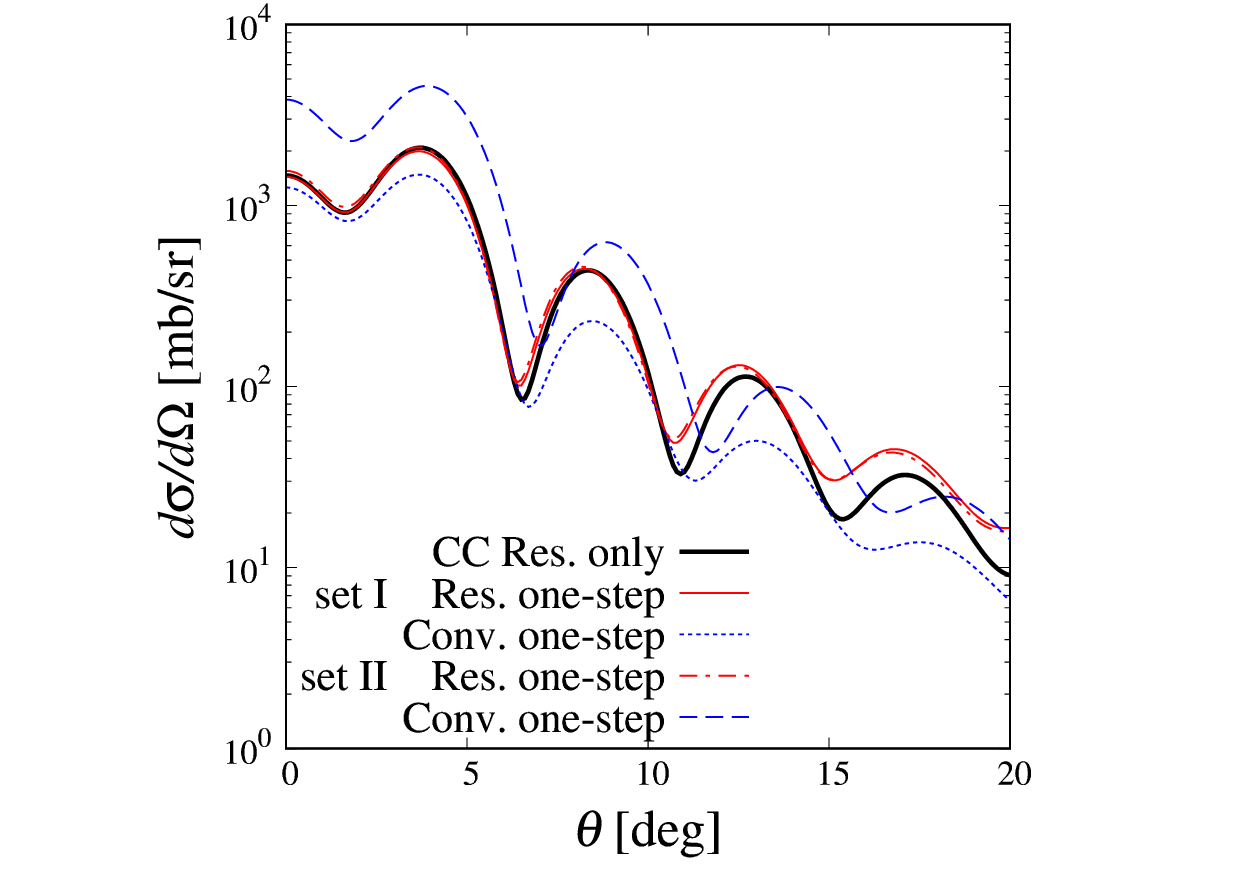}
 \caption{
 Inelastic cross sections for the $^6$Li + $^{40}$Ca reaction at $E$ = 26 MeV/nucleon
 obtained from the one-step calculations with Eqs. \eqref{eq:1step-eq} 
 and \eqref{eq:1step-eq-res2}.
 The pseudo states considered in the results are calculated with
 parameter sets I or II for Gaussian bases. 
 }
 \label{fig:6Li-setI-vs-setII}
\end{figure}

In Fig.~\ref{fig:6Li-setI-vs-setII}, we consider the model-space
dependence by comparing the inelastic cross sections obtained from the
conventional and resonant one-step calculations. 
The thick solid line denotes the result of the CDCC with only fragmented resonant
states for set I, which is the same as the dot-dashed line in Fig.~\ref{fig:6Li-inela}.
The thin solid and dotted
lines represent the results obtained for set~I from the conventional and
resonant one-step calculations, respectively. The difference between the
two results originates from coupling effects between fragmented resonant
states. 
The dot-dashed and dashed lines show the results
obtained for set~II. One can see that the result of the resonant
one-step calculation for set~I is in good agreement with that for
set~II. Furthermore the result of the resonant one-step calculation is
also consistent with that of the CDCC with only fragmented resonant
states. This means that the multi-step effects between the ground and
resonant states are negligible.
On the other hand, the results obtained from the conventional one-step
calculation vary with the model space, with substantially higher values
for set II. This problem is considered to originate from the omission of
coupling effects between the fragmented resonant states in the
conventional one-step calculations. Thus, we conclude that the resonant
one-step calculation should be adopted to estimate the multi-step
effects for the resonant state in a manner that is independent of 
the size of the model space in the CDCC framework.
\begin{figure}[t]
 \begin{center}
  \includegraphics[scale=0.75]
  {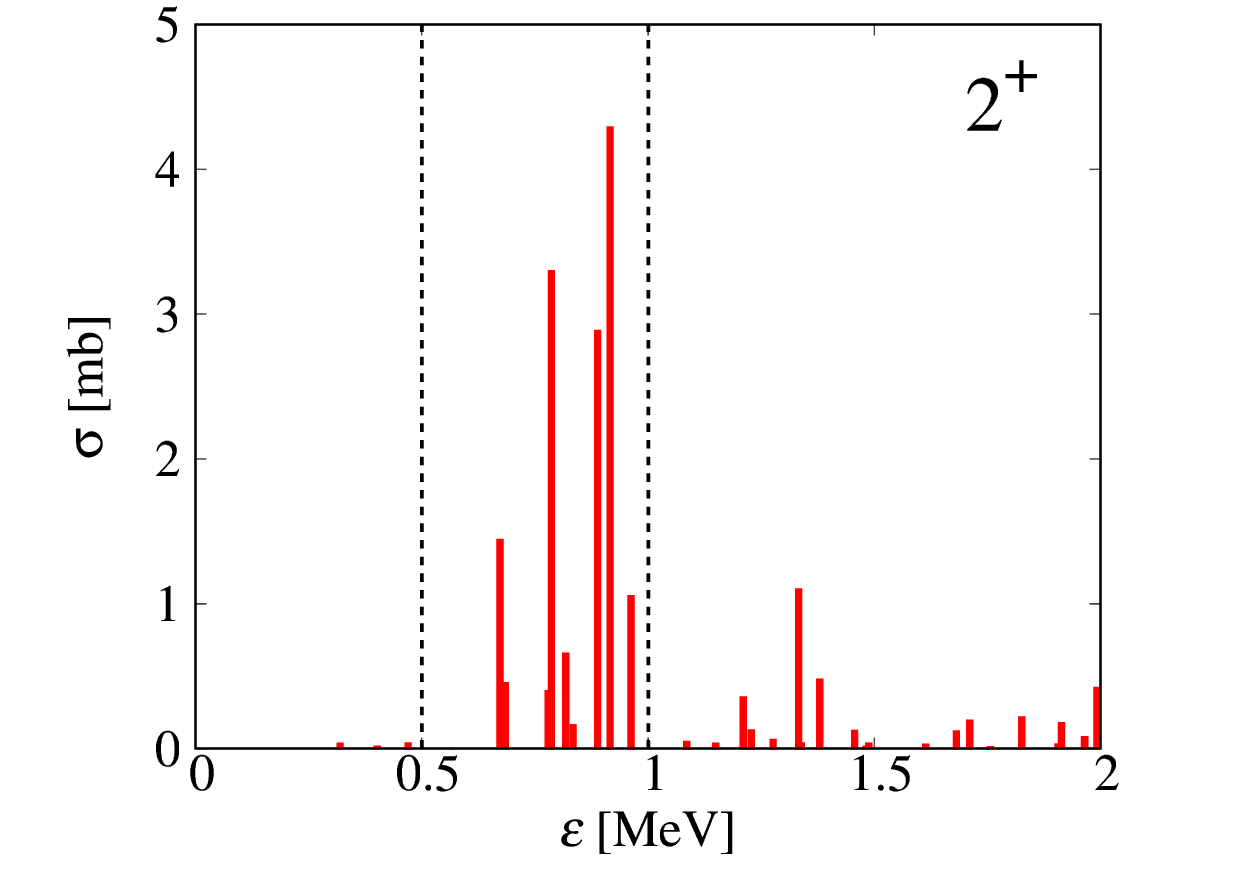}   
  \caption{
  Breakup cross section for the $^6$He + $p$ reaction 
  at $E$ = 41 MeV/nucleon describing the transition to 
  the $2^+$ continuum states obtained from the CDCC.
  States in the area between the two lines represent fragmented resonant
  states. 
  }
  \label{fig:6He-dbu}
 \end{center}  
\end{figure}
%
\subsection{$^6$He scattering}

\begin{figure}[t]
 \begin{center}
  \includegraphics[scale=0.8]{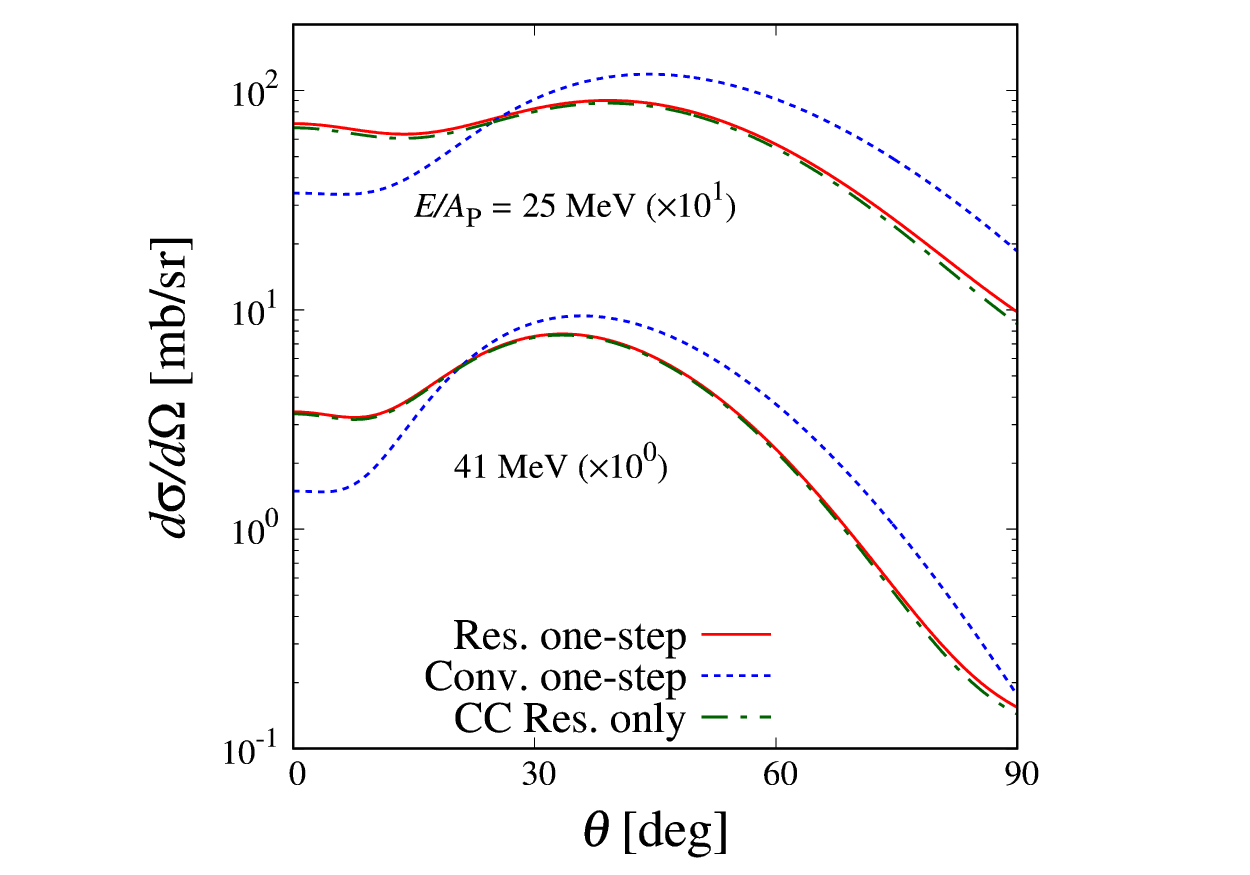}
  \caption{
  Inelastic cross sections for the $^6$He + $p$ reaction at $E$ = 41 and
  25 MeV/nucleon obtained from the CDCC including only the fragmented resonant
  states and one-step calculations using Eqs. \eqref{eq:1step-eq} 
  and \eqref{eq:1step-eq-res2}.
  }
  \label{fig:6He-inela}
 \end{center}  
\end{figure}
\begin{figure}[t]
 \begin{center}
  \includegraphics[scale=0.8]{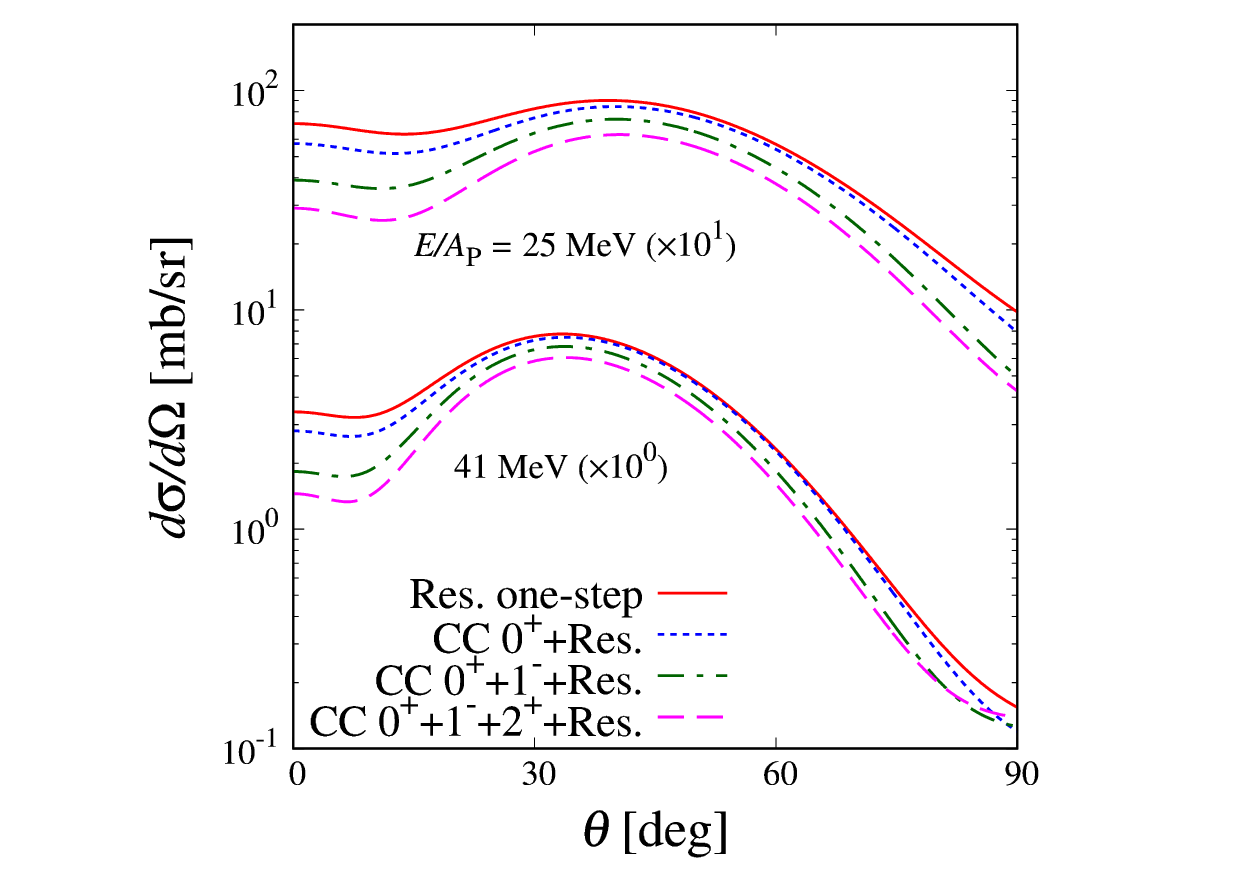}
  \caption{
  Inelastic cross sections for the $^6$He + $p$ reaction at $E$ = 41 and
  25 MeV/nucleon obtained from the CDCC
  and one-step calculations using Eq.~\eqref{eq:1step-eq-res2}.
  }
  \label{fig:6He-cc-dwba}
 \end{center}  
\end{figure}

Finally, we discuss the multi-step effects in the $^6$He + $p$ reactions
at $E$ = 41 and 25 MeV/nucleon, in which $^6$He is described by the $^4$He +
$n$ + $n$ three-body model. 
In this analysis, we use the same model parameters as described in 
Ref.~\cite{Oga20}. The resonant energy and decay width of the $2^+_1$
state are 0.848 MeV and 0.136 MeV, respectively. In contrast to the
two-body model of $^6$Li, the discretized breakup cross sections in the
three-body model of $^6$He exhibit no regularity with respect to the
internal energy of $^6$He, as shown in Fig.~\ref{fig:6He-dbu}. 
Therefore, it is not easy to identify the fragmented resonant states. 
In the present calculations, we select states within the range of
$0.5$~MeV~$\leq~\varepsilon\leq$~$1.0$~MeV as the fragmented resonant
states for simplicity. The number of the fragmented states is ten.  
Recently, a new method has been proposed for
characterizing resonant states out of pseudo states obtained by
diagonalizing a three-body Hamiltonian \cite{Cas19}. Application of this
technique to the present study will be interesting and an
important future work.

In Fig.~\ref{fig:6He-inela},
we show the inelastic cross sections to the fragmented resonant
$2^+_1$states calculated using Eq.~\eqref{eq:inela}. 
It should be noted that the dashed line is consistent with the cross
section to the $2^+_1$ state calculated with the complex-scaling method
shown in Fig.~3 in Ref.~\cite{Oga20}.
The solid and dot-dashed lines indicate the results of
the resonant one-step calculation and the CDCC with only the fragmented
resonant states, respectively.
One can see that the solid line is in good agreement
with the dot-dashed line.
This indicates that the multi-step effects between the ground state
and $2^{+}_{1}$ state are negligible,
which is consistent with the results of the AMD-MCC reported in Ref.~\cite{Kan20-2}.
On the other hand, the value from the conventional one-step calculation
shown by the dotted line is larger than one from the resonant one-step
calculation at backward angles, and it looks like the multi-step effects
are much significant. However, the result of the conventional one-step
calculation depends on the size of three-body ($^4$He + $n$ + $n$) model
space as mentioned above, and  
the difference between the dotted and dot-dashed lines no
longer indicates the multi-step effects.
We thus conclude that the multi-step effects between the ground state
and $2^+_1$ state shown by the difference between the solid and
dot-dashed lines are negligible. 

Furthermore, we discussed the coupling effects on the resonant state
in Fig.~\ref{fig:6He-cc-dwba}.
The solid line indicates the result of the resonant one-step calculation.
The dotted (dot-dashed) line represents the result of the CDCC with coupling between
only $I^\pi$ = $0^{+}$ ($0^{+}$, $1^{-}$) non-resonant states and 
the fragmented resonant states.
The dashed line corresponds to the result of the CDCC with coupling between
all states in the model space.
It is found that the cross sections decrease as the coupling effects on the non-resonant
states are taken.
These effects reduce the cross section by approximately 20--30\% at 41 MeV/nucleon and
become stronger as the incident energy decreases.

\section{Summary}
We have investigated the multi-step effects between the ground and resonant states
by comparing the results of CC calculations and one-step calculations.
In the CDCC, the resonant state is fragmented into multiple discretized states,
where the number of fragmented resonant states is dependent on the size of the model space.
In this work, we introduced two approaches for the one-step calculations,
which we refer to as conventional one-step calculation and resonant one-step calculation.
The former includes no coupling effects between the fragmented resonant states,
whereas the latter considers only the multi-step effects between the fragmented resonant 
states.

In the analysis of $^6$Li + $^{40}$Ca scattering,
we confirmed the model-space dependence in the conventional and resonant 
one-step calculations.
The numerical results obtained from the conventional one-step calculation did not converge
with increasing the size of the model space; in other words, this result was unphysical.
In contrast, the results of the resonant one-step calculation were independent of the size 
of the model space.
Therefore, the resonant one-step calculation is suitable for investigating the multi-step 
effects between resonant states and other states.
We have also discussed the multi-step effects for the resonance $2^{+}_{1}$
in $^6$He via the $^6$He + $p$ reaction.
The numerical results demonstrated that the multi-step effects between the $2^{+}_{1}$ and
ground states were negligible in accordance with Ref. \cite{Kan20-2},
in which the resonance $2^{+}_{1}$ was obtained as a single state.
We thus conclude that it is reasonable to describe the resonant state as an excited bound 
state.
Meanwhile, the coupling effects to non-resonant states for $I^\pi=0^+$,
$1^-$, and $2^+$ are important and reduce the cross section by
approximately 20--30\% at 41 MeV/nucleon and become stronger at 25 MeV/nucleon. 
This result originates from the fact that $^6$He is an unstable nucleus.
Therefore, analyses of unstable nuclei should take into account the coupling effects to 
non-resonant states.

\section*{Acknowledgments}
This work is supported in part by Grant-in-Aid for Scientific Research
(No.\ JP18K03650, 18K03617, and 18H05407) from Japan Society for the
Promotion of Science (JSPS).

\appendix
\section*{Appendix}

\begin{figure*}[h]
 \begin{center}
  \includegraphics[scale=1.3]{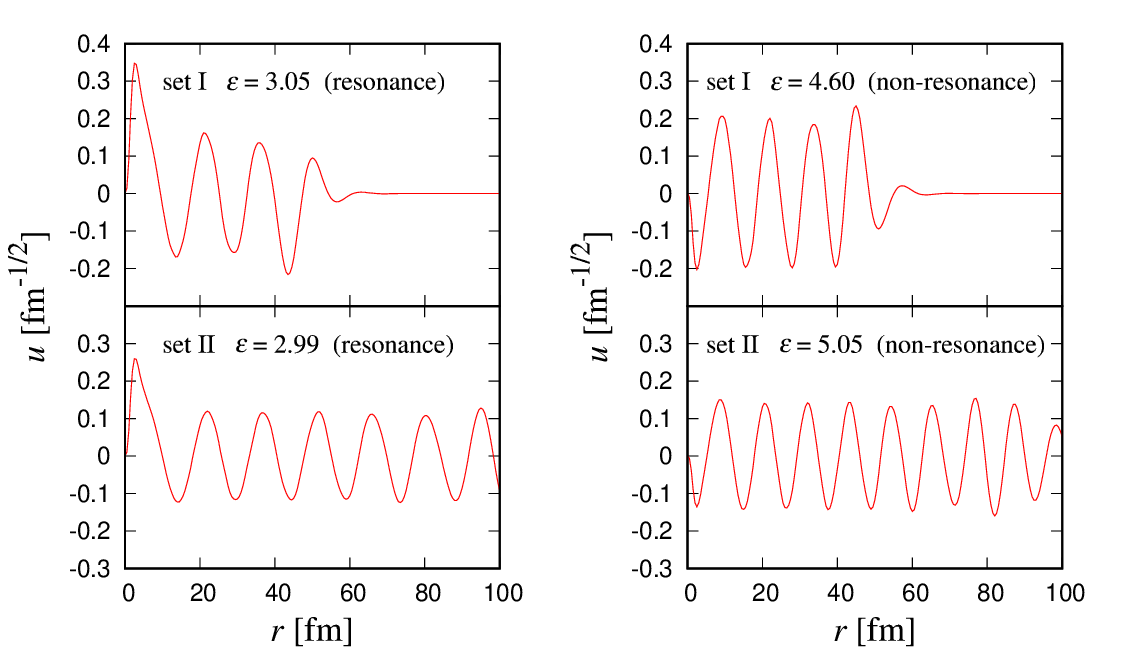}
  \caption{
  The radial wavefunctions of $^6$Li in sets I and II.
  }
  \label{fig:wf-setI-setII}
 \end{center}  
\end{figure*}

In Fig.~\ref{fig:wf-setI-setII}, we show the radial wavefunctions $u$ of the fragmented 
resonant and non-resonant states, which are obtained by diagonalizing the internal Hamiltonian
of $^6$Li in sets I and II.
The wavefunctions are normalized as $\int_{0}^{\infty}|u|^{2}dr = 1$.
As the fragmented resonant states,
we select states with $\varepsilon = 3.05$ MeV in set I (left top panel) and 
2.99 MeV in set II (left bottom panel),
which are near the resonant energy of 2.75 MeV.
The states with $\varepsilon = 4.60$ MeV in set I (right top panel) and 
$\varepsilon = 5.05$ MeV in set II (right bottom panel) correspond to
the non-resonant states.
One can see the wavefunctions in set II oscillate at a larger distance than
those in set I because the maximum value of the range parameter $r_{\rm max}$ in
Gaussian bases for sets I and II are taken as 20 fm and 40 fm, respectively.

\begin{figure*}[h]
 \begin{center}
  \includegraphics[scale=1.3]{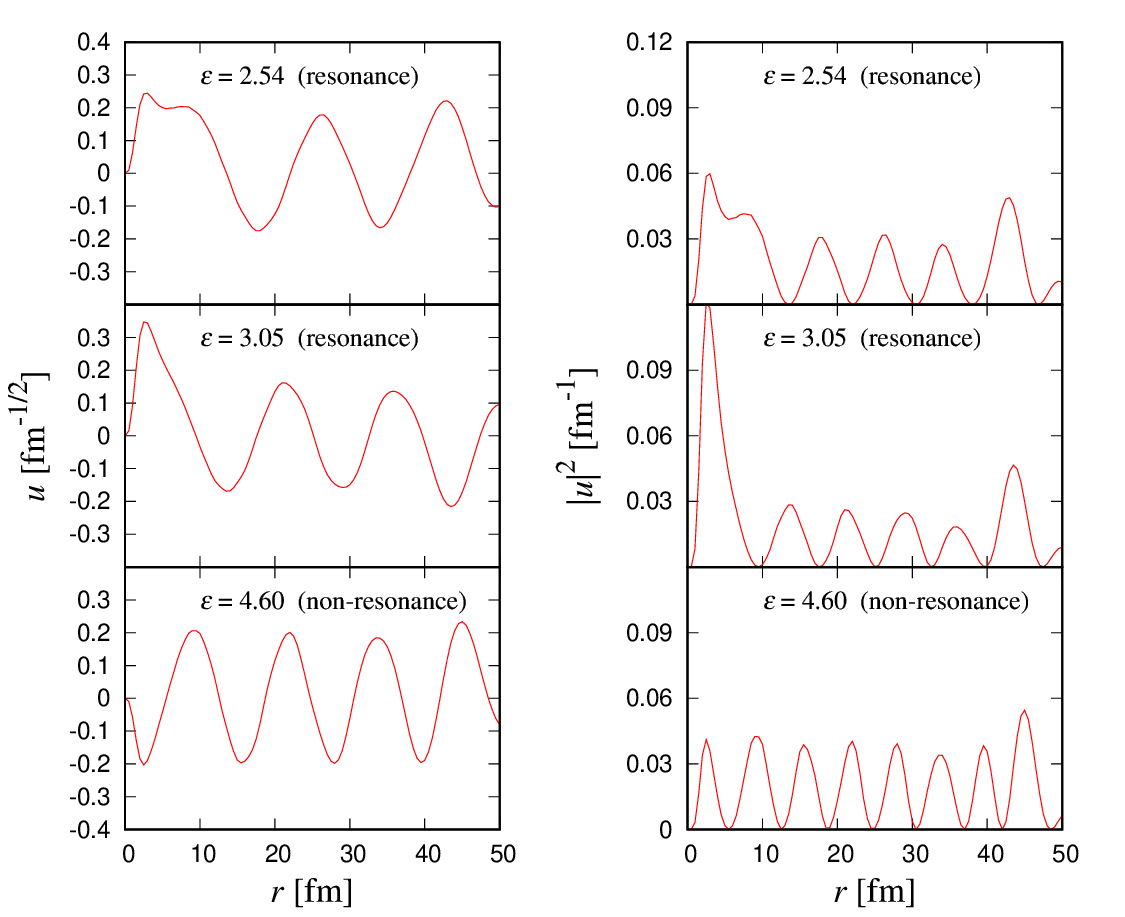}
  \caption{
  The radial wavefunctions of $^6$Li and the probability densities in set I.
  }
  \label{fig:wf-setI}
 \end{center}  
\end{figure*}

Figure~\ref{fig:wf-setI} shows the radial wavefunctions in set I.
The left and right panels represent the wavefunctions and the probability densities,
respectively.
Here we select states with $\varepsilon = 2.54$ MeV (top panel) and 3.05 MeV (mid panel)
as the resonant state,
and a state with $\varepsilon = 4.60$ MeV (bottom panel) as the non-resonant state.
It is found that the fragmented resonant states have large amplitudes in the internal 
region up to $r_{\rm max}$.
On the other hand,
the non-resonant state has a constant amplitude in the whole region.
To clear this point, 
we show the probabilities obtained by integrating the probability densities over $r$ up to
$r_{\rm max}$ in Table~\ref{tab:integral},
and the values of the fragmented resonant states are larger than 
one of the non-resonant state.
\begin{table}[htbp]
 \tblcaption{
 The probabilities obtained of the fragmented resonant and non-resonant states in set I
 by integrating the probability densities over $r$ up to 20 fm.
 }
 \label{tab:integral}
 \begin{tabular}
  {cp{4em} cp{10em} cp{10em} cp{10em} cp{10em} cp{10em} cp{10em} cp{10em}}
  \hline
  &\multicolumn{1}{c}{~}
  &\multicolumn{1}{c}{~~2.54 MeV}
  &\multicolumn{1}{c}{~~~~3.05 MeV}
  &\multicolumn{1}{c}{~~~~4.06 MeV}
  \\
  &\multicolumn{1}{c}{~~}
  &\multicolumn{1}{c}{~resonance}
  &\multicolumn{1}{c}{~~~resonance}
  &\multicolumn{1}{c}{~~~non-resonance}
  \\  
  \hline
  &\multicolumn{1}{c}{Probability}
  &\multicolumn{1}{c}{~~0.532}
  &\multicolumn{1}{c}{~~~~0.548}
  &\multicolumn{1}{c}{~~~~0.369}
  \\
  \hline
 \end{tabular}       
\end{table}


\end{document}